\documentclass{article}

\usepackage{PRIMEarxiv}

\usepackage[utf8]{inputenc} % allow utf-8 input
\usepackage[T1]{fontenc}    % use 8-bit T1 fonts
\usepackage{hyperref}       % hyperlinks
\usepackage{url}            % simple URL typesetting
\usepackage{booktabs}       % professional-quality tables
\usepackage{amsfonts}       % blackboard math symbols
\usepackage{nicefrac}       % compact symbols for 1/2, etc.
\usepackage{microtype}      % microtypography
\usepackage{lipsum}
\usepackage{fancyhdr}       % header
\usepackage{graphicx}       % graphics
\graphicspath{{media/}}     % organize your images and other figures under media/ folder
\usepackage{xspace}
\usepackage{amsmath}

\newcommand{\etal}{\emph{et al.}\xspace}

\makeatletter
\newcommand{\thickhline}{%
  \noalign{\global\savedwidth\arrayrulewidth\global\arrayrulewidth 1pt}%
  \hline
  \noalign{\global\arrayrulewidth\savedwidth}%
}
\newlength{\savedwidth}
\makeatother

\title{MRI-to-CT synthesis using drifting models}

\author{
  Qing Lyu, Jeremy Hudson, Chirstopher T. Whitlow \\
  Department of Radiology and Biomedical Imaging \\
  Yale School of Medicine \\
  New Haven, CT\\
  \texttt{\{qing.lyu, jeremy.hudson, christopher.whitlow\}@yale.edu} \\
   \And
  Jianxu Wang, Ge Wang \\
  Department of Biomedical Engineering \\
  Rensselaer Polytechnic Institute \\
  Troy, NY\\
  \texttt{\{wangj68, wangg6\}@email} \\
}

\begin{document}
\maketitle

\begin{abstract}
Accurate MRI-to-CT synthesis could enable MR-only pelvic workflows by providing CT-like images with bone details while avoiding additional ionizing radiation. In this work, we investigate recently proposed drifting models for synthesizing pelvis CT images from MRI and benchmark them against convolutional neural networks (UNet, VAE), a generative adversarial network (WGAN-GP), a physics-inspired probabilistic model (PPFM), and diffusion-based methods (FastDDPM, DDIM, DDPM). Experiments are performed on two complementary datasets: Gold Atlas Male Pelvis and the SynthRAD2023 pelvis subset. Image fidelity and structural consistency are evaluated with SSIM, PSNR, and RMSE, complemented by qualitative assessment of anatomically critical regions such as cortical bone and pelvic soft-tissue interfaces. Across both datasets, the proposed drifting model achieves high SSIM and PSNR and low RMSE, surpassing strong diffusion baselines and conventional CNN-, VAE-, GAN-, and PPFM-based methods. Visual inspection shows sharper cortical bone edges, improved depiction of sacral and femoral head geometry, and reduced artifacts or over-smoothing, particularly at bone–air–soft tissue boundaries. Moreover, the drifting model attains these gains with one-step inference and inference times on the order of milliseconds, yielding a more favorable accuracy–efficiency trade-off than iterative diffusion sampling while remaining competitive in image quality. These findings suggest that drifting models are a promising direction for fast, high-quality pelvic synthetic CT generation from MRI and warrant further investigation for downstream applications such as MRI-only radiotherapy planning and PET/MR attenuation correction. 
\end{abstract}

% keywords can be removed

\section{Introduction}
Medical imaging often requires both Computed Tomography (CT) and Magenetic Resonance Imaging (MRI) modalities for comprehensive clinical assessment\textemdash CT provides Hounsfield unit (HU) electron density information essential for radiotherapy dose calculation, while MRI offers superior soft-tissue contrast for tumor delineation and diagnosis. Acquiring both modalities increases cost, radiation exposure, and introduces misalignment between scans \cite{pelc2014recent, liguori2015emerging, brenner2007computed}. In addition, CT acquisition may be constrained by practical factors such as equipment availability and maintenance burden, leading to variability in access and leaving some clinical or research cohorts without CT coverage \cite{frija2021improve, burdorf2022comparing}. These limitations, together with the relative rarity of high-quality paired and well-registered CT–MRI datasets, make CT synthesis from MRI clinically important because it can provide CT-like quantitative surrogates for CT-dependent tasks (e.g., attenuation correction and MRI-only radiotherapy workflows) when CT is unavailable or undesirable \cite{siam2025paired, edmund2017review}

Cross-modality synthesis—generating one modality from the other—has been a major research direction, but existing approaches face notable limitations \cite{edmund2017review, bahloul2024advancements, dayarathna2024deep, ibrahim2025generative}. Early strategies include atlas-based registration and patch-based learning, while more recent deep learning solutions span supervised convolutional neural networks (e.g., U-Net variants) and generative modeling frameworks \cite{guerreiro2017evaluation, uh2014mri, burgos2015ct, chen2017attenuation, roy2014mr, lee2017multi}. Generative approaches such as VAEs and GANs can improve realism and sharpness but may suffer from blurring (VAEs) or training instability and hallucinated structures (GANs), especially near bone–air–soft tissue interfaces \cite{iyer2025mri, Li_2025, skandarani2023gans, lei2019mri, wang2022magnetic, hsu2022synthetic, lemus2022dosimetric, yang2020unsupervised, matsuo2022unsupervised, gong2024channel, liu2021ct}. Diffusion models have shown strong fidelity and robustness, yet their iterative sampling can impose substantial inference-time costs that limit clinical practicality, motivating interest in faster generative alternatives \cite{pan2024synthetic, ozbey2023unsupervised, lyu2022conversion}.

Drifting models are a recent family of generative methods that learn a continuous “drift” (vector-field) dynamics to transform a simple source distribution into the target image distribution, making them naturally suited to conditional synthesis tasks such as MRI-to-CT by learning how an input-conditioned representation should evolve toward a CT-like output \cite{deng2026generative}. Compared with diffusion models, drifting models can be implemented with fewer integration steps and therefore offer faster inference in practice while targeting comparable perceptual fidelity. Relative to GAN-based synthesis, which may be sensitive to training instability and can produce visually plausible yet unfaithful structures, drifting models provide an appealing balance between realism and controllability for anatomy-preserving medical translation.

In this study, we investigate recently emerged drifting models for pelvic MRI-to-CT synthesis and compare them against representative baselines including convolutional neural network (CNN), variational autoencoder (VAE), GAN, and diffusion models. We conduct experiments on two datasets\textemdash SynthRAD2023 and Gold Atlas Male Pelvis\textemdash and evaluate performance using structural similarity index (SSIM) and peak signal-to-noise ratio (PSNR), alongside qualitative assessment in anatomically critical regions. Our key contributions are: (1) modification of original drifting models and adding MRI as conditions for controllable CT synthesis, (2) a systematic benchmark of drifting models against established synthesis families on two complementary pelvic datasets, and (3) evidence that drifting models can achieve high-quality synthetic CT generation with notably fast inference, offering a favorable accuracy–efficiency trade-off for downstream use.

\section{Related Studies}
\label{sec:headings}

\subsection{Atlas/registration-based CT synthesis approaches}
Early sCT pipelines typically rely on aligning one or more atlas MR–CT pairs to a target MRI and then fusing the warped CTs into a pseudo-CT; these methods are conceptually appealing because they encode prior anatomical correspondences, but they can be sensitive to registration quality and population mismatch. Uh \etal is a representative early study in this vein, demonstrating the feasibility of MRI-based radiotherapy treatment planning by generating pseudo-CT via atlas registration and then assessing its suitability for dose calculation \cite{uh2014mri}. Extending the atlas concept toward PET/MR workflows, Burgos \etal focus on head-and-neck attenuation correction and describe an iterative multi-atlas strategy, underscoring that reliable pseudo-CT generation is also critical when CT is unavailable for attenuation mapping in PET/MR \cite{burgos2015ct}. Building directly on multi-atlas synthesis for radiotherapy, Guerreiro \etal evaluate a multi-atlas sCT approach for MRI-only treatment planning in head-and-neck and prostate cases, comparing it to a bulk-density assignment alternative and showing that multi-atlas synthesis can achieve clinically small dose differences while improving automatic bone delineation \cite{guerreiro2017evaluation}. In parallel, Roy \etal highlight a complementary perspective: rather than treating synthesis only as the final deliverable, they use image synthesis to enable improved MR-to-CT brain registration, illustrating how synthesis can reduce multimodal mismatch during alignment \cite{roy2014mr}. Lee \etal further refine the atlas pipeline by combining multi-atlas fusion with patch-based refinement, aiming to correct local details that global registration-and-fusion may miss \cite{lee2017multi}.

\subsection{VAE/GAN-based approaches}
As CT synthesis shifted from atlas/registration pipelines to generative modeling, VAE- and GAN-based approaches became prominent because they enable direct MRI-to-CT image translation with substantially faster inference and improved representation power. In broad terms, VAE-based methods emphasize stable optimization and anatomically faithful reconstructions through probabilistic latent representations, whereas GAN-based methods tend to deliver sharper, more realistic-looking CT textures but often require additional constraints to mitigate hallucinations and preserve geometry in unpaired or weakly paired training.

On the VAE side, Iyer \etal propose an MRI-to-CT synthesis framework for pediatric cranial imaging built on a VAE. They further incorporate cranial suture segmentations as an auxiliary supervised signal, and report high structural similarity for sCT along with strong skull and suture segmentation performance on an in-house pediatric dataset \cite{iyer2025mri}.

On the GAN side, much of the progress is driven by adding explicit mechanisms to improve structural reliability under imperfect pairing. Lei \etal introduce dense cycle-consistent GANs for MRI-only sCT generation, using cycle-consistency to reduce dependence on strictly paired datasets while encouraging anatomically plausible translation \cite{lei2019mri}. Yang \etal extend this line with a structure-constrained CycleGAN by adding a structure-consistency loss based on the modality-independent neighborhood descriptor, explicitly targeting geometric distortions that can arise in unconstrained CycleGAN mappings \cite{yang2020unsupervised}. For head-and-neck radiotherapy, Liu \etal present a multi-cycle GAN that strengthens cyclic constraints to stabilize learning and better preserve anatomy in this challenging region \cite{liu2021ct}. In thoracic imaging, Matsuo \etal develop an unsupervised attention-based framework for chest MRI to CT translation, reflecting the need for stronger guidance when motion and low-signal regions make structure preservation difficult \cite{matsuo2022unsupervised}. Gong \etal further push constraint design with cycleSimulationGAN, combining channel-wise attention with structural-similarity constraints\textemdash including a contour mutual information loss\textemdash to improve detail recovery and structural retention for unpaired head-and-neck MR-to-CT synthesis \cite{gong2024channel}.

Several works also emphasize clinical readiness and evaluation beyond image similarity. Wang \etal apply GAN-based MRI-to-CT generation for intracranial tumor radiotherapy planning, illustrating a clinically oriented deployment focused on planning feasibility \cite{wang2022magnetic}. Hsu \etal study sCT generation for MRI-guided adaptive radiotherapy in prostate cancer, where models must produce reliable sCT repeatedly as anatomy evolves over the treatment course \cite{hsu2022synthetic}. Lemus \etal reinforce that clinical acceptance should be assessed with task-based endpoints—such as dose metrics, DVH agreement, and gamma analysis—rather than relying solely on pixel-wise similarity \cite{lemus2022dosimetric}. Finally, Skandarani \etal provide an empirical study of GANs for medical image synthesis, offering practical insight into GAN behavior and common evaluation pitfalls that are important when interpreting and comparing GAN-based sCT pipelines \cite{skandarani2023gans}.

\subsection{Diffusion-based approaches}
Diffusion-based synthesis replaces one-shot generation with a learned conditional denoising process, which often improves fidelity and reduces some GAN-specific failure modes at the cost of iterative sampling. Lyu and Wang explores MRI-to-CT conversion using diffusion and score-matching models, representing an early effort to apply diffusion-based models to CT synthesis \cite{lyu2022conversion}. Pan \etal presents an approach using a 3D transformer-based denoising diffusion model with a Gaussian forward noise process on CT and a reverse denoising process conditioned on MRI, reporting improved image metrics and dosimetric agreement with generation in minutes \cite{pan2024synthetic}. Özbey \etal proposes unsupervised medical image translation with adversarial diffusion models, using a conditional diffusion translation mechanism plus adversarial/cycle-style constraints to enable unpaired cross-modal translation while accelerating sampling through larger reverse steps \cite{ozbey2023unsupervised}. Li \etal proposes FDDM, an unsupervised MR→CT translation method that separates anatomy conversion from appearance generation and uses frequency-decoupled diffusion with dual reverse paths for low/high frequency components, reporting improved fidelity and anatomical accuracy on brain and pelvis MR→CT translation benchmarks \cite{Li_2025}. 

\section{Methodology}

\subsection{Dataset}
\subsubsection{Gold Atlas Male Pelvis Dataset}
The Gold Atlas Male Pelvis Dataset is a public pelvic MRI-CT dataset comprising 19 male patients, where each case includes a planning CT, a T1-weighted MRI, a T2-weighted MRI, and a deformably registered CT volume \cite{nyholm2018goldatlas}. The data were acquired at three Swedish radiotherapy departments using site-specific clinical protocols and different scanners, which makes the dataset suitable for validating cross-site MRI-to-CT synthesis methods. For CT, the reported scanners were Siemens Somatom Definition AS+, Toshiba Aquilion, and Siemens Emotion 6, with slice thicknesses of 3.0, 2.0, and 2.5 $mm$, respectively, and in-plane pixel sizes of approximately 0.98 $\times$ 0.98, 1.0 $\times$ 1.0, and 0.98 $\times$ 0.98 $mm^2$. For MRI, the dataset includes both T1-weighted and T2-weighted acquisitions; the T1-weighted scans were acquired with slice thicknesses of 3.0, 2.0, and 2.5 $mm$ and in-plane pixel sizes ranging from 0.875 $\times$ 0.875 to 1.1 $\times$ 1.1 $mm^2$, while the T2-weighted scans used 2.5 $mm$ slice thickness and in-plane pixel sizes from 0.875 $\times$ 0.875 to 1.1 $\times$ 1.1 $mm^2$. The original dataset article reports site-specific acquisition settings rather than a single common image matrix for all subjects, reflecting the heterogeneous multi-center design.

\subsubsection{SynthRAD2023 Dataset}
SynthRAD2023 is a large multi-center benchmark for synthetic CT generation in radiotherapy, containing 540 paired MRI-CT sets and 540 paired cone beam CT-CT sets across brain and pelvis anatomies \cite{thummerer2023synthrad}. In this work, we use the MRI-to-CT pelvis subset (Task 1 pelvis), which contains 270 paired MRI-CT scans. This subset was collected from two centers: Center A acquired pelvic MRI scans on Philips Ingenia 1.5T/3.0T systems using 3D T1-weighted spoiled gradient echo imaging, with in-plane pixel spacing between 0.94 and 1.14 $mm$, rows ranging from 400 to 528, and columns ranging from 103 to 528; Center C acquired scans on Siemens MAGNETOM Avanto, Skyra, and Vida 3T systems using 3D T2-weighted SPACE imaging, with in-plane pixel spacing between 1.17 and 1.30 $mm$ and matrix size 288 $\times$ 384. The paired CT scans were reconstructed at 512 $\times$ 512 in-plane size, with pixel spacing of 0.77-1.37 $mm$ at Center A and 0.98-1.17 $mm$ at Center C, and slice thicknesses of 1.5 to 3.0 $mm$ and 2.0 to 3.0 $mm$, respectively.

\subsection{Image preprocessing}
All MRI and CT volumes were first resampled to an isotropic spatial resolution of 1 $\times$ 1 $\times$ 1 $mm^3$. After resampling, each slice was centrally cropped or zero padded to ensure a uniform in-plane image size of 512 $\times$ 512. This step standardizes the field of view across subjects and simplifies mini-batch training. Finally, intensity normalization was performed independently for each scan so that inter-subject intensity scale variations were reduced before network training. Let $x$ denote an input volume and $T(\cdot)$ the preprocessing operator; then the final network input can be written as
\begin{equation}
x_{\mathrm{prep}} = T(x),
\label{eq1}
\end{equation}
where $T(\cdot)$ includes isotropic resampling, central cropping or zero padding, and per-scan intensity normalization.

\subsection{Drifting Model}
Drifting models reformulate generative modeling as a train-time distribution evolution problem rather than an iterative test-time denoising process. Instead of progressively refining a noisy sample during inference, the model learns during optimization how the generator distribution should move toward the target data distribution. As a result, the transport process is absorbed into training, while inference remains a one-step mapping from MRI to synthetic CT \cite{deng2026generative}.

Let $m$ denote an MRI input, $c$ the corresponding target CT, and $\epsilon \sim p(\epsilon)$ random noise. A conditional generator $f_{\theta}$ produces a synthetic CT $\hat{c}$ based on conditional image $m$ and noise $\epsilon$
\begin{equation}
\hat{c} = f_{\theta}(m,\epsilon).
\label{eq2}
\end{equation}
In the spirit of drifting models, adapted from the paper’s unconditional pushforward formulation, the proposed conditional output distribution $q_{\theta}(\cdot \mid m)$ is expressed as follows
\begin{equation}
q_{\theta}(\cdot \mid m) = \left[f_{\theta}(m,\cdot)\right]_{\#} p(\epsilon).
\label{eq3}
\end{equation}

As suggested for the original drifting model paper, we explore the drifting loss in both the image domain and the feature space. For the image domain, Let $\hat{c}$ denote a generated CT image, let $c^{+}$ denote a real CT sample from the target distribution, and let $c^{-}$ denote a generated CT sample from the current model distribution. The drifting field is therefore defined directly on image intensities, so the generator is trained to move each synthesized CT toward nearby real CT images while repelling it from other generated CT images. This follows the original raw-space drifting formulation, but is adapted here to conditional MRI-to-CT synthesis.

For a generated sample \(\hat{c}\), the general drifting field is written as
\begin{equation}
\mathbf{V}_{p,q}(\hat{c}) = \mathbb{E}_{c^{+}\sim p(\cdot \mid m)} \mathbb{E}_{c^{-}\sim q(\cdot \mid m)} \left[ \mathcal{K}(\hat{c},c^{+},c^{-}) \right].
\label{eq4}
\end{equation}
where $\mathcal{K}(\cdot,\cdot,\cdot)$ is a kernel-like function describing interactions among three sample points.

The drift field in drifting models is built from attraction toward data samples and repulsion from generated samples, with anti-symmetry ensuring zero drift at equilibrium when the two distributions match:
\begin{equation}
\mathbf{V}^{+}_{p}(\hat{c}) = \frac{1}{Z_{p}(\hat{c})} \mathbb{E}_{c^{+}\sim p(\cdot \mid m)} \left[ k(\hat{c},c^{+})(c^{+}-\hat{c}) \right],
\label{eq5}
\end{equation}

\begin{equation}
\mathbf{V}^{-}_{q}(\hat{c}) = \frac{1}{Z_{q}(\hat{c})} \mathbb{E}_{c^{-}\sim q(\cdot \mid m)} \left[ k(\hat{c},c^{-})(c^{-}-\hat{c}) \right],
\label{eq6}
\end{equation}
where the normalization factors are
\begin{equation}
Z_{p}(\hat{c}) = \mathbb{E}_{c^{+}\sim p(\cdot \mid m)} \left[ k(\hat{c},c^{+}) \right], \qquad Z_{q}(\hat{c}) = \mathbb{E}_{c^{-}\sim q(\cdot \mid m)} \left[ k(\hat{c},c^{-}) \right].
\label{eq7}
\end{equation}
$k(\cdot, \cdot)$ is the pairwise similarity kernel
\begin{equation}
k(a, b) = \exp(-\frac{1}{\tau}\left\|a-b\right\|_2).
\label{eq8}
\end{equation}

The final drifting field is defined as
\begin{equation}
\mathbf{V}_{p,q}(\hat{c}) = \mathbf{V}^{+}_{p}(\hat{c}) - \mathbf{V}^{-}_{q}(\hat{c}).
\label{eq9}
\end{equation}

At equilibrium state, the generator output should satisfy the fixed-point relation
\begin{equation}
f_{\hat{\theta}}(m,\epsilon) = f_{\hat{\theta}}(m,\epsilon) + \mathbf{V}_{p,q_{\hat{\theta}}}\!\left(f_{\hat{\theta}}(m,\epsilon)\right).
\label{eq10}
\end{equation}

This motivates the training-time update
\begin{equation}
f_{\theta_{i+1}}(m,\epsilon) \leftarrow f_{\theta_i}(m,\epsilon) + \mathbf{V}_{p,q_{\theta_i}}\!\left(f_{\theta_i}(m,\epsilon)\right).
\label{eq11}
\end{equation}

We convert this update rule into the drifting loss
\begin{equation}
\mathcal{L}_{\mathrm{drift}} = \mathbb{E}_{m,\epsilon} \left[ \left\| f_{\theta}(m,\epsilon) - \mathrm{sg}\!\left( f_{\theta}(m,\epsilon) + \mathbf{V}_{p,q_{\theta}}\!\left(f_{\theta}(m,\epsilon)\right) \right) \right\|_2^2 \right],
\label{eq12}
\end{equation}
where \(\mathrm{sg}(\cdot)\) denotes the stop-gradient operator.

For paired MRI-to-CT synthesis, an additional voxel-level reconstruction term can be included to preserve anatomical correspondence:
\begin{equation}
\mathcal{L}_{\mathrm{total}} = \lambda_{\mathrm{drift}}\mathcal{L}_{\mathrm{drift}} + \lambda_{1}\left\| \hat{c}-c\right\|_{1},
\label{eq13}
\end{equation}
where $\lambda_{\mathrm{drift}}$ and $\lambda_{1}$ control the balance between distribution-level drifting and voxel-wise fidelity.

For the feature domain drifting loss computation, we first pre-train a ResNet-18 model $\phi$ on real CT images and then freeze all model parameters for feature extraction. For a generated CT image $\hat{c}$, a real CT sample $c^+$, and a generated CT sample $c^-$, define their feature embeddings as $z=\phi(\hat{c})$, $z^+=\phi(c^+)$, $z^-=\phi(c^-)$ respectively. Then, we modify \eqref{eq4} to \eqref{eq7} and \eqref{eq9} so that the general drifting field is written in feature space

\begin{equation}
\mathbf{V}^{\phi}_{p,q}(\hat{z}) = \mathbb{E}_{z^{+}\sim p^{\phi}(\cdot \mid m)} \mathbb{E}_{z^{-}\sim q^{\phi}_{\theta}(\cdot \mid m)} \left[\mathcal{K}(\hat{z},z^{+},z^{-})\right],
\label{eq14}
\end{equation}

\begin{equation}
\mathbf{V}^{+,\phi}_{p}(\hat{z}) = \frac{1}{Z^{\phi}_{p}(\hat{z})}\mathbb{E}_{z^{+}\sim p^{\phi}(\cdot \mid m)}\left[k_{\phi}(\hat{z},z^{+})(z^{+}-\hat{z})\right],
\label{eq15}
\end{equation}

\begin{equation}
\mathbf{V}^{-,\phi}_{q}(\hat{z}) = \frac{1}{Z^{\phi}_{q}(\hat{z})}\mathbb{E}_{z^{-}\sim q^{\phi}_{\theta}(\cdot \mid m)}\left[k_{\phi}(\hat{z},z^{-})(z^{-}-\hat{z})\right],
\label{eq16}
\end{equation}

\begin{equation}
Z^{\phi}_{p}(\hat{z}) = \mathbb{E}_{z^{+}\sim p^{\phi}(\cdot \mid m)}\left[k_{\phi}(\hat{z},z^{+})
\right], \qquad
Z^{\phi}_{q}(\hat{z}) = \mathbb{E}_{z^{-}\sim q^{\phi}_{\theta}(\cdot \mid m)}\left[k_{\phi}(\hat{z},z^{-})\right],
\label{eq17}
\end{equation}

\begin{equation}
\mathbf{V}^{\phi}_{p,q}(\hat{z}) = \mathbf{V}^{+,\phi}_{p}(\hat{z})-\mathbf{V}^{-,\phi}_{q}(\hat{z}).
\label{eq18}
\end{equation}

\subsection{Implementation Details}
We split both Gold Atlas and SynthRAD2023 datasets based on 70\%/10\%/20\% rule to create training, validation, and testing subsets, respectively. The ResNet-18 model is pre-trained on real CT images in the training subset using a self-supervised contrastive loss for 300 epochs, following \cite{chen2020simple}. The conditional generation in the drifting model is a UNet-like model. For drifting loss computation, 16 positive and negative samples are created by randomly cropping 64 $\times$ 64 patches from real image and generated images, respectively. In \eqref{eq13}, $\lambda_{drift}$ and $\lambda_{1}$ are empirically set to 1 and 10, respectively. Adam optimizer is used and the learning rate is set to 0.0001. The training loss is recorded and monitored during the training process. When the training loss does not decrease more than 1\% for the most recent 20 epochs, the early stop triggers and the training model is saved for the later inference. All experiments are conducted using Pytorch via a single Nvidia H200 GPU.

\section{Results}
To ensure a fair comparison, we benchmark the proposed MRI‑conditioned drifting model against a diverse set of representative baselines, including convolutional encoder–decoder architectures (UNet, VAE) \cite{2015arXiv150504597R, 2013arXiv13126114K}, an adversarial model (WGAN‑GP) \cite{gulrajani2017improved}, a physics‑inspired probabilistic framework (PPFM) \cite{2024ITRPM8788H}, and several diffusion‑based approaches (FastDDPM, DDIM, DDPM) \cite{2024arXiv240514802J, 2020arXiv201002502S, 2020arXiv200611239H}. All models are implemented with a comparable number of trainable parameters and similar backbone capacity, so that performance differences primarily reflect the underlying generative mechanism and conditioning strategy rather than model size. DDPM inference is based on a 1000-step denoising, while DDIM and fastDDPM is implemented via a 20-step sampling. PPFM inference results are made by a 18-step sampling process.

\subsection{Results on Gold Atlas Male Pelvis dataset}
\begin{figure}[ht]
   \centering
   \includegraphics[width=\textwidth]{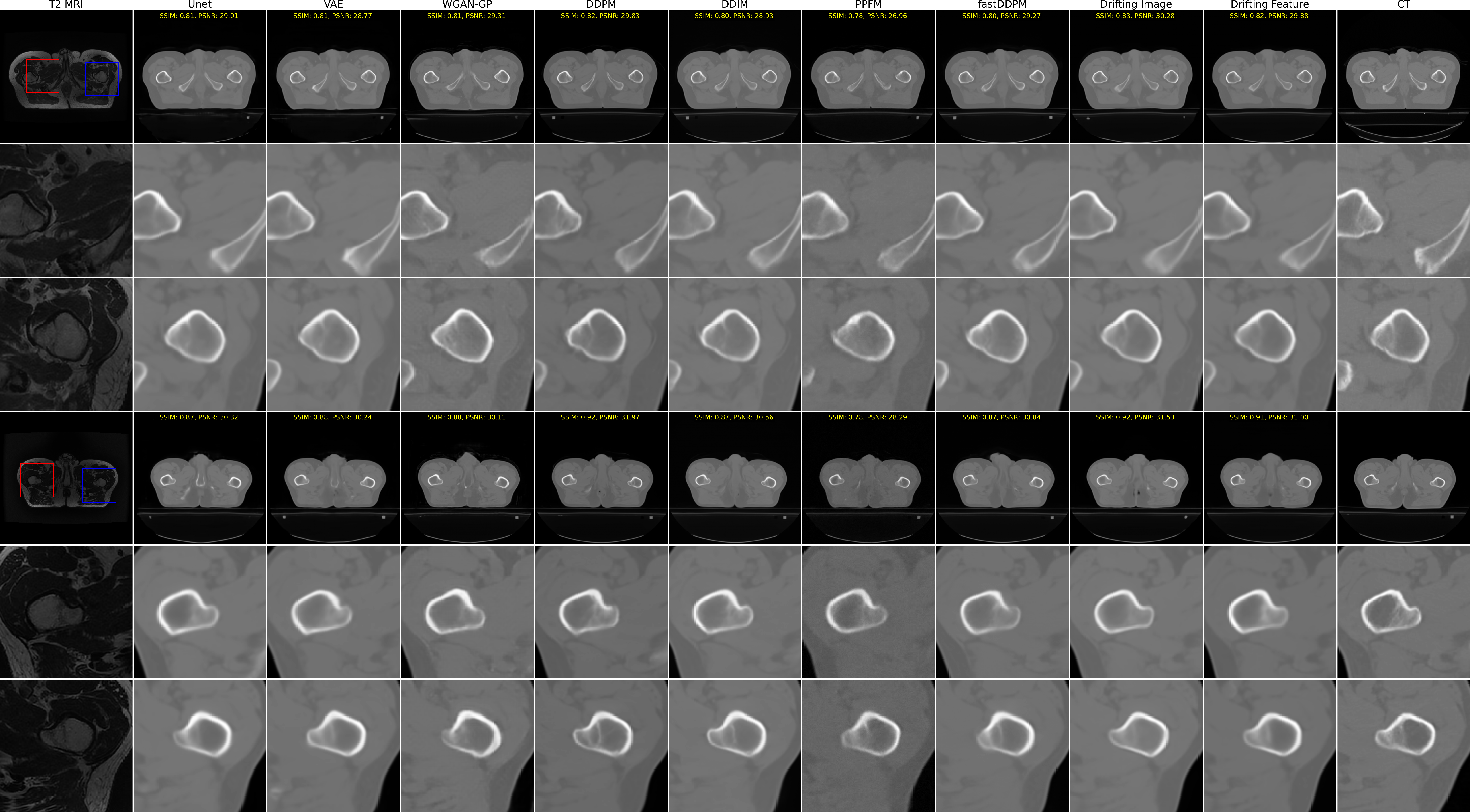}
   \caption{Comparison of CT synthesize results from different methods on Gold Atlas Dataset.}
   \label{fig_1} 
\end{figure}

On the Gold Atlas dataset, the proposed drifting model outperforms all competing methods except DDPM in terms of SSIM, PSNR, and RMSE. Quantitatively, it attains high SSIM and PSNR and low RMSE, surpassing the diffusion-based models such as FastDDPM, PPFM, and DDIM, as well as non‑diffusion methods such as UNet, VAE, and WGAN‑GP. The only exception is DDPM, which obtains the highest SSIM and PSNR and the lowest RMSE. Visually, Figure \ref{fig_1} shows that the proposed drifting model better preserves fine pelvic structures and soft‑tissue boundaries, with clearer organ details and reduced over‑smoothing compared with UNet/VAE and fewer noise‑induced artifacts than WGAN‑GP and PPFM. Compared with other diffusion models, drifting produces sharper cortical bone edges and more accurate Hounsfield unit transitions, which align more closely with the reference CT. Drifting models on image space and feature space obtain comparative results with slightly higher SSIM and PSNR values for the drifting model on image space.

A representative example in Figure \ref{fig_1} illustrates these trends: UNet and VAE yield overly smooth bone–soft‑tissue transitions, WGAN‑GP introduces artifacts that hinder image quality, and DDIM/PPFM/fastDDPM produce good results with clear tissue boundary but lower image quality when comparing with DDPM. Among all approaches, DDPM demonstrates the best ability to clearly demonstrate bone details and tissue boundary. As for the proposed drifting model, it can bring about comparable results with DDIM/fastDDPM and present clear tissue boundary and bone structure. Table \ref{table1} presented quantitative comparison of synthetic CT from different approaches. It can be found that the proposed drifting model slightly overcome DDIM and fastDDPM but not as good as DDPM.

\begin{table}
 \caption{Quantitative comparison of various methods}
  \centering
  \begin{tabular}{c | c c c | c c c}
    \thickhline
    & \multicolumn{3}{c|}{Gold Atlas} & \multicolumn{3}{c}{SynthRAD2023} \\
    \cline{2-7}
    & SSIM & PSNR & RMSE & SSIM & PSNR & RMSE \\
    \hline
    UNet & 0.835 $\pm$ 0.030 & 29.53 $\pm$ 1.44 & 0.041 $\pm$ 0.003 & 0.935 $\pm$ 0.012 & 28.74 $\pm$ 1.68 & 0.040 $\pm$ 0.017   \\
    VAE & 0.840 $\pm$ 0.039 & 29.26 $\pm$ 1.74 & 0.043 $\pm$ 0.003 & 0.931 $\pm$ 0.013 & 29.29 $\pm$ 1.87 & 0.043 $\pm$ 0.012   \\
    WGAN-GP & 0.853 $\pm$ 0.035 & 29.64 $\pm$ 1.01 & 0.045 $\pm$ 0.002 & 0.909 $\pm$ 0.007 & 27.68 $\pm$ 1.69 & 0.045 $\pm$ 0.021   \\
    PPFM & 0.823 $\pm$ 0.054 & 27.08 $\pm$ 2.25 & 0.084 $\pm$ 0.019 & 0.871 $\pm$ 0.028 & 27.18 $\pm$ 2.05 & 0.048 $\pm$ 0.023   \\
    FastDDPM & 0.865 $\pm$ 0.041 & 30.58 $\pm$ 1.38 & 0.039 $\pm$ 0.003 & 0.948 $\pm$ 0.015 & 29.83 $\pm$ 1.75 & 0.033 $\pm$ 0.010   \\
    DDIM & 0.854 $\pm$ 0.042 & 30.13 $\pm$ 1.42 & 0.042 $\pm$ 0.004 & 0.946 $\pm$ 0.021 & 30.12 $\pm$ 1.64 & 0.035 $\pm$ 0.013   \\
    DDPM & \textbf{0.889 $\pm$ 0.045} & \textbf{31.34 $\pm$ 1.64} & \textbf{0.031 $\pm$ 0.003} & \textbf{0.969 $\pm$ 0.012} & \textbf{31.51 $\pm$ 1.84} & \textbf{0.024 $\pm$ 0.008}  \\
    Drifting (image) & 0.869 $\pm$ 0.043 & 30.56 $\pm$ 1.23 & 0.035 $\pm$ 0.002 & 0.948 $\pm$ 0.009 & 30.82 $\pm$ 1.69 & 0.030 $\pm$ 0.011   \\
    Drifting (feature) & 0.863 $\pm$ 0.040 & 30.37 $\pm$ 1.25 & 0.036 $\pm$ 0.002 & 0.955 $\pm$ 0.010 & 30.91 $\pm$ 1.65 & 0.030 $\pm$ 0.011   \\
    \thickhline
  \end{tabular}
  \label{table1}
\end{table}

\subsection{Inference speed comparison}
\begin{figure}[ht]
   \centering
   \includegraphics[width=\textwidth]{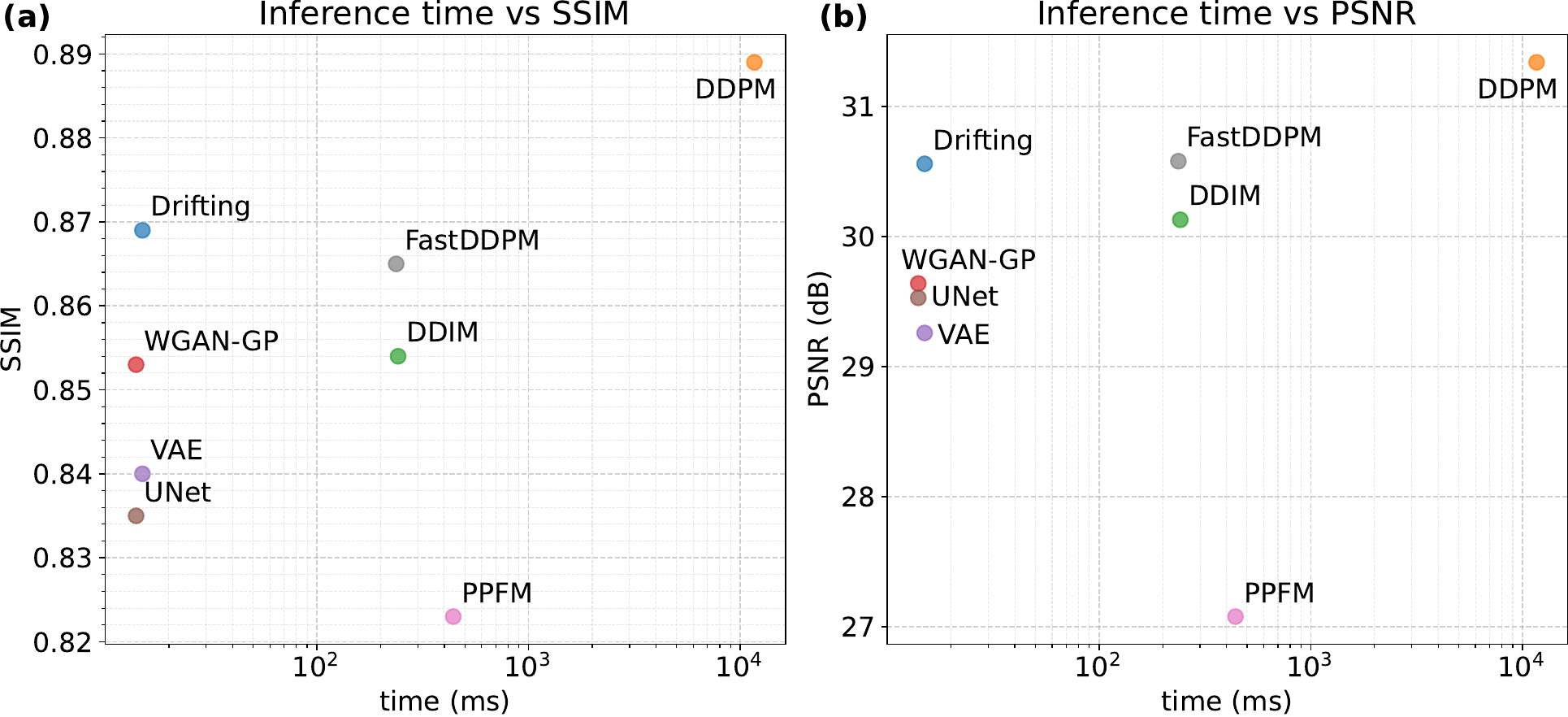}
   \caption{Comparison of result image quality versus inference time.}
   \label{fig_2} 
\end{figure}

Figure \ref{fig_2} summarizes the trade‑off between inference time and image quality for all competing methods on the Gold Atlas dataset. The proposed drifting model achieves one of the lowest inference times (on the order of  ms) while simultaneously obtaining the highest SSIM and PSNR, forming the most favorable position in both “time vs SSIM” and “time vs PSNR” plots. DDPM reaches competitive SSIM/PSNR but requires longer sampling, shifting it toward slower inference, whereas DDIM and FastDDPM reduce runtime at the cost of noticeable drops in SSIM and PSNR. Conventional generative baselines such as VAE and WGAN‑GP do not close the quality gap to drifting despite similar or higher computational cost.

% These results indicate that conditioning the drifting model on MRI not only improves reconstruction fidelity but also enables efficient sampling with relatively few denoising steps. Consequently, the method offers a more favorable quality–efficiency trade‑off than existing diffusion and non‑diffusion approaches, which is important for time‑critical radiotherapy planning workflows.

\subsection{Results on SynthRAD2023 dataset}
\begin{figure}[ht]
   \centering
   \includegraphics[width=\textwidth]{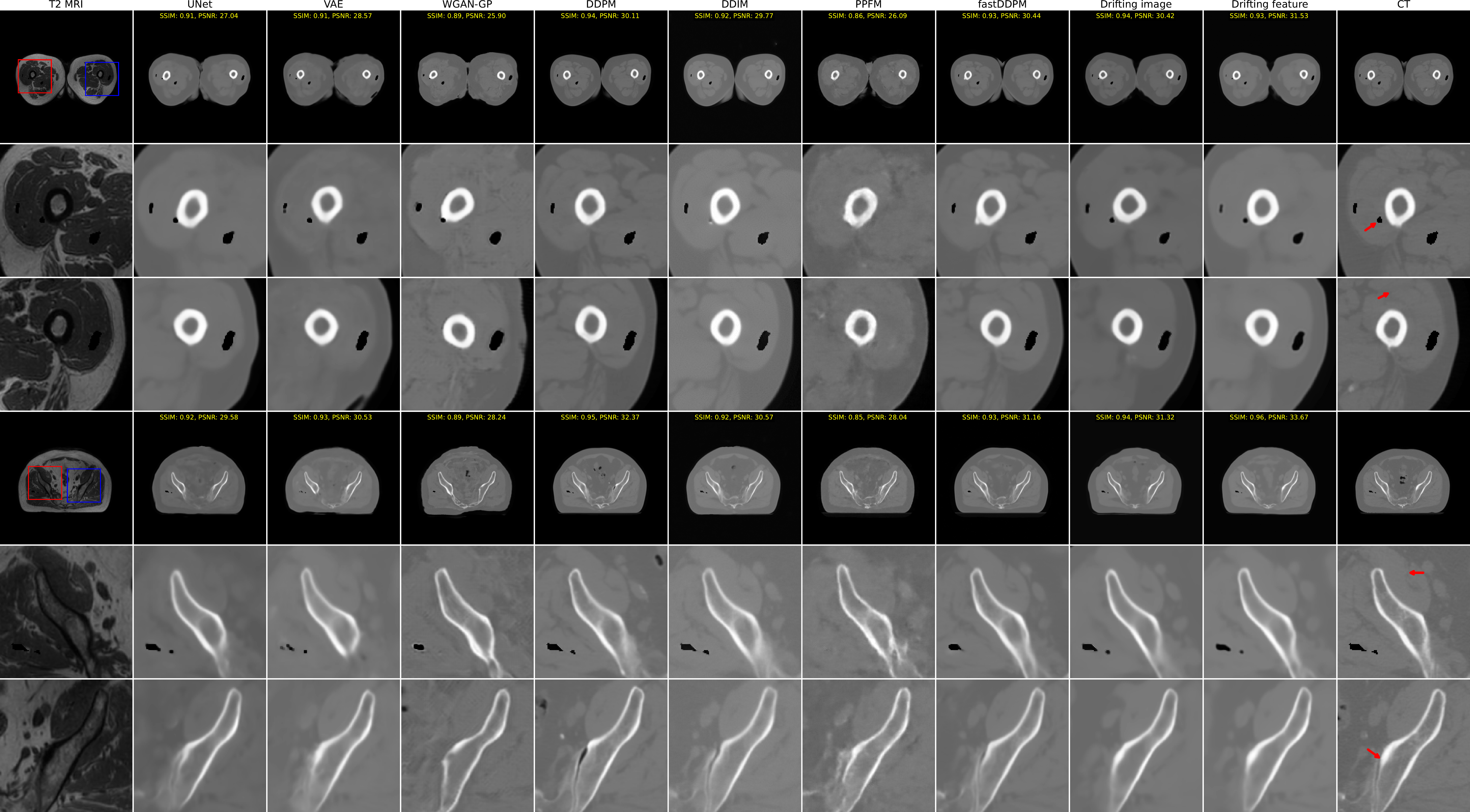}
   \caption{Comparison of CT synthesize results from different methods on SynthRAD2023 Dataset.}
   \label{fig_3} 
\end{figure}

On the more diverse SynthRAD2023 dataset, the proposed drifting model again yields superior quantitative performance, achieving high SSIM and PSNR with low RMSE. Compared to UNet, VAE, WGAN‑GP, and PPFM, drifting substantially improves structural fidelity and intensity accuracy, and it further improves over DDIM and FastDDPM despite their strong baseline performance. Qualitative examples in Figure \ref{fig_3} show that drifting better captures pelvic bone geometry and soft‑tissue contrast under varying anatomies and acquisition conditions, whereas competing methods either over‑smooth high‑contrast regions or introduce spurious texture in low‑dose areas. Drifting models on image space and feature space also obtain comparative results. Differing from results on Gold Atlas datset, results on SynthRAD2023 show slightly better results for the drifting model on feature space.

Figure \ref{fig_3} also demonstrates that the drifting model maintains robust performance across different patients, with consistent depiction of sacral curvature, femoral heads, and rectal filling patterns that closely match the ground‑truth CT. Even in challenging cases with atypical positioning or metallic implants, drifting introduces fewer streaks and unnatural intensity transitions than GAN‑based and other diffusion baselines, supporting its generalizability beyond the controlled Gold Atlas cohort.

\subsection{Investigation on model uncertainty}
\begin{figure}[ht]
   \centering
   \includegraphics[width=\textwidth]{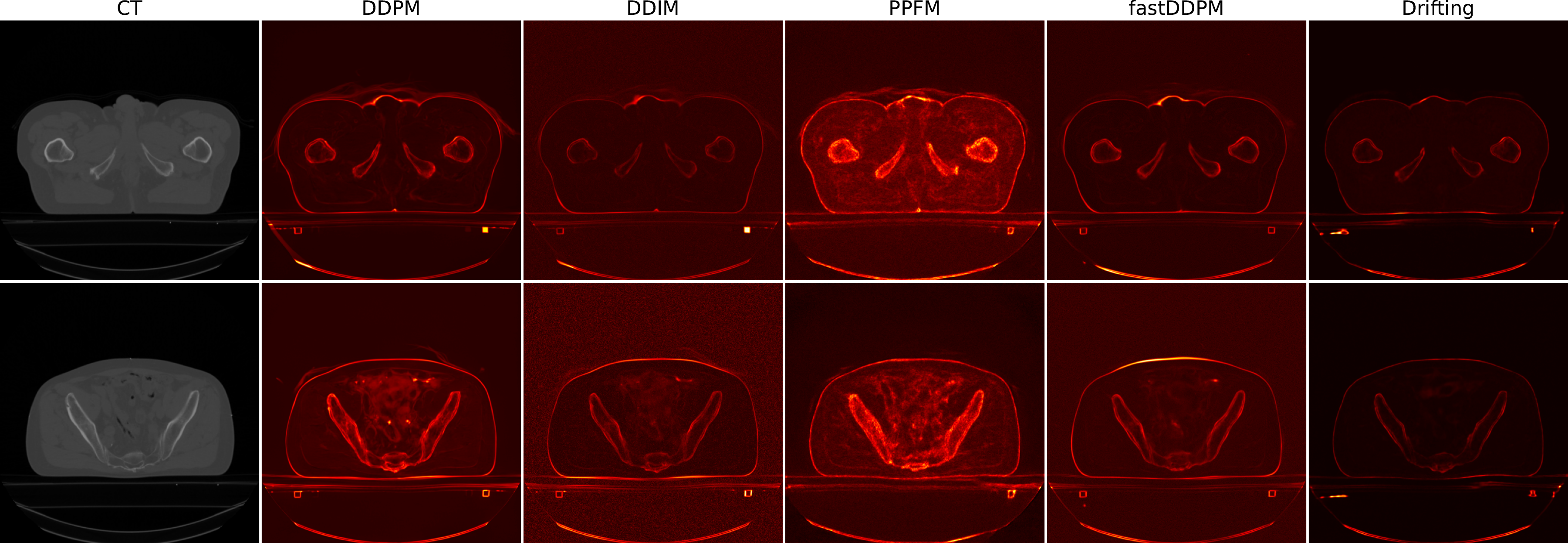}
   \caption{Pixel-vise standard deviation map based on 20 sampling results for each comparing approach.}
   \label{fig_4} 
\end{figure}

Although inference diversity can be adjusted by method-specific sampling parameters such as temperature in \eqref{eq8} for the drifting model or $\sigma_t$ in DDPM-based methods, we believe that uncertainty investigation remains necessary. A controllable diversity parameter does not by itself explain how sensitive a model is to stochastic perturbations under a fixed inference configuration, whereas repeated-sampling analysis directly measures the variability of the generated result for the same MRI input. Therefore, in this study, we evaluate epistemic uncertainty using pixel-wise standard deviation maps obtained from 20 repeated samplings with different random noises, as shown in Figure \ref{fig_4}. Under this unified protocol, the standard deviation map serves as a practical indicator of sampling stability: a lower and more spatially confined variance implies that the model is less sensitive to random initialization and thus more reliable at inference time.

Figure \ref{fig_4} presents qualitative comparisons of model uncertainty for two representative cases. Across the compared methods, the standard deviation maps reveal clear differences in the spatial distribution. PPFM exhibits broader highlighted regions with a relatively elevated global uncertainty level, indicating stronger sensitivity to stochastic sampling, whereas DDPM shows a comparatively more constrained pattern but still retains noticeable uncertainty in several structures. In comparison, the proposed drifting model shows a distinct uncertainty pattern relative to the baseline methods. Its standard deviation map highlights fewer or more localized regions, suggesting improved control of sampling variability and more stable inference behavior under repeated random initialization.

\section{Discussion}
In this work, we systematically evaluated a conditional drifting model for pelvic MRI-to-CT synthesis on two complementary public datasets, Gold Atlas and SynthRAD2023, and compared it against convolutional, variational, adversarial, and diffusion-based baselines. Across both cohorts, the drifting model consistently achieved competitive or superior image quality as measured by SSIM, PSNR, and RMSE, while also delivering fast inference times on the order of milliseconds per slice. These results indicate that drifting-based generative modeling provides a favorable balance between fidelity and computational efficiency, which is particularly relevant for time-sensitive radiotherapy workflows where multiple synthetic CT volumes may need to be generated repeatedly over a treatment course.

An important observation from our experiments is that the drifting model preserves pelvic bone geometry and soft-tissue boundaries more faithfully than CNN, VAE, and GAN baselines, which either tended toward over-smoothing or introduced artifacts in high-contrast regions. On Gold Atlas, qualitative inspection showed sharper cortical bone edges, clearer rectal wall delineation, and fewer checkerboard or noise-like patterns than WGAN-GP and PPFM, aligning with the superior SSIM/PSNR and lower RMSE reported for the drifting model. On the more heterogeneous SynthRAD2023 pelvis subset, the drifting model remained robust across patients and acquisition protocols, maintaining accurate depiction of sacrum curvature, femoral heads, and rectal filling patterns even in challenging cases with atypical positioning or metallic implants. Together, these findings suggest that the attraction–repulsion mechanism in drifting models can improve structural consistency at anatomy-critical interfaces compared with purely voxel-wise or adversarial objectives.

Compared with diffusion-based methods, the proposed approach occupies an interesting point in the design space. Prior diffusion models for MRI-to-CT synthesis have demonstrated high fidelity and strong dosimetric agreement, but at the cost of iterative sampling that can require seconds to minutes per volume. In our study, DDPM achieved strong SSIM and PSNR but incurred substantially higher inference time, whereas accelerated variants such as DDIM and FastDDPM narrowed the runtime gap at the expense of noticeable drops in image quality. By contrast, the drifting model internalizes the transport dynamics into training, allowing one-step inference conditioned on MRI while still matching or surpassing diffusion baselines in both similarity metrics and qualitative sharpness. This accuracy–efficiency trade-off suggests that drifting models may serve as a pragmatic alternative when clinical throughput is a primary constraint, for example in MRI-only planning or adaptive radiotherapy scenarios where same-session sCT generation is desirable.

Our results should also be interpreted in the context of broader work on MRI-to-CT synthesis and generative medical imaging. Traditional atlas-based and registration-driven methods encode anatomical priors but can be brittle to registration errors and population mismatch, particularly in multi-center settings such as SynthRAD2023. VAE- and GAN-based frameworks have improved realism and flexibility but often require sophisticated structural constraints and careful training to avoid blurring or hallucinated structures. Recent diffusion approaches mitigate some GAN failure modes and provide strong baselines, but practicality can be an issue for integration into routine clinical pipelines. The drifting framework, operating directly in image space with a kernel-based attraction–repulsion field, offers an alternative that is simple to condition on MRI, yields deterministic one-step mapping at test time, and empirically reduces the tendency toward over-smoothing or texture artifacts around bone–air–soft tissue interfaces. These characteristics make drifting models a promising candidate to complement, rather than replace, existing generative paradigms in synthetic CT research.

Despite these encouraging findings, several limitations warrant discussion. First, our experiments are restricted to pelvic MRI-to-CT synthesis and rely solely on image similarity metrics (SSIM, PSNR, RMSE) and qualitative anatomical assessment, without explicit evaluation on downstream clinical tasks such as dose calculation, DVH agreement, or attenuation correction performance. Prior work has emphasized that task-based endpoints are critical for judging clinical readiness; thus, future studies should integrate radiotherapy planning and PET/MR attenuation experiments to quantify the dosimetric and functional impact of drifting-based sCT. Second, we focused on paired supervised training and did not explore unpaired or semi-supervised regimes, which are highly relevant given the scarcity of well-registered MRI–CT pairs and the availability of large unpaired datasets. Extending drifting models to unpaired translation, potentially by combining them with cycle or structural consistency constraints, could further broaden their applicability. Third, our implementation uses a relatively standard UNet-like backbone and a simple image-domain drifting field with a fixed Gaussian kernel and globally chosen loss weights. It is plausible that more expressive architectures (e.g., transformer or hybrid CNN–transformer backbones) or adaptive, spatially varying kernels could further enhance performance, particularly in regions with complex anatomy or large intensity heterogeneity. Additionally, we did not perform an extensive hyperparameter search across all baselines; while we attempted to match model capacity and training protocols, some performance differences may partly reflect implementation details or optimization choices. Finally, all experiments were conducted on a single high-end GPU, and we did not systematically analyze memory consumption, scalability to 3D or 4D (time-resolved) volumes, or behavior under domain shifts such as different scanners, sequences, or institutions beyond those included in the two public datasets.

Future work will therefore include several directions. From a methodological perspective, incorporating anatomy-aware or frequency-decoupled losses into the drifting field, as well as exploring multi-modal conditioning (e.g., joint T1/T2, or MRI plus prior segmentations), may improve fine-detail reconstruction and robustness across varying acquisition protocols. From an evaluation standpoint, we plan to conduct comprehensive task-based studies in radiotherapy planning and PET/MR attenuation, including comparisons of dose metrics, gamma analysis, and clinical acceptability ratings between drifting-based sCT and ground-truth CT. Finally, extending the proposed framework beyond the pelvis to other anatomies such as brain, head-and-neck, and thorax, and assessing its performance under domain adaptation scenarios, will be essential to understand how well drifting models generalize in realistic multi-center clinical settings.

\section{Conclusion}
We presented a conditional drifting model for pelvic MRI-to-CT synthesis and benchmarked it against convolutional, variational, adversarial, and diffusion-based approaches on the Gold Atlas Male Pelvis and SynthRAD2023 pelvis datasets. Across both cohorts, the drifting model achieved consistently high SSIM and PSNR together with low RMSE, while preserving fine pelvic bone structures and soft-tissue boundaries with fewer artifacts than competing methods. Compared with diffusion-based baselines, the proposed approach offers a more favorable accuracy–efficiency trade-off, attaining comparable or superior image quality with substantially reduced inference time through one-step generation. These properties make drifting models attractive for integration into time-critical MR-only workflows, including radiotherapy planning and attenuation correction, where reliable and fast synthetic CT generation is required. Future work will extend this framework to additional anatomies and task-based evaluations, and investigate unpaired training and architecture refinements to further enhance robustness and clinical readiness.

%Bibliography
\bibliographystyle{unsrt}  
\bibliography{references}

@article{pelc2014recent,
  title={Recent and future directions in CT imaging},
  author={Pelc, Norbert J},
  journal={Annals of biomedical engineering},
  volume={42},
  number={2},
  pages={260--268},
  year={2014},
  publisher={Springer}
}

@article{liguori2015emerging,
  title={Emerging clinical applications of computed tomography},
  author={Liguori, Carlo and Frauenfelder, Giulia and Massaroni, Carlo and Saccomandi, Paola and Giurazza, Francesco and Pitocco, Francesca and Marano, Riccardo and Schena, Emiliano},
  journal={Medical Devices: Evidence and Research},
  pages={265--278},
  year={2015},
  publisher={Taylor \& Francis}
}

@article{brenner2007computed,
  title={Computed tomography—an increasing source of radiation exposure},
  author={Brenner, David J and Hall, Eric J},
  journal={New England journal of medicine},
  volume={357},
  number={22},
  pages={2277--2284},
  year={2007},
  publisher={Mass Medical Soc}
}

@article{frija2021improve,
  title={How to improve access to medical imaging in low-and middle-income countries?},
  author={Frija, Guy and Bla{\v{z}}i{\'c}, Ivana and Frush, Donald P and Hierath, Monika and Kawooya, Michael and Donoso-Bach, Lluis and Brklja{\v{c}}i{\'c}, Boris},
  journal={EClinicalMedicine},
  volume={38},
  year={2021},
  publisher={Elsevier}
}

@article{burdorf2022comparing,
  title={Comparing magnetic resonance imaging and computed tomography machine accessibility among urban and rural county hospitals},
  author={Burdorf, Benjamin T},
  journal={Journal of Public Health Research},
  volume={11},
  number={1},
  pages={jphr--2021},
  year={2022},
  publisher={SAGE Publications Sage UK: London, England}
}

@article{siam2025paired,
  title={A paired CT and MRI dataset for advanced medical imaging applications},
  author={Siam, Zakaria Shams and Akon, Md Younus and Munmun, Israt Jahan and Al-Amin, Abdullah and Salam, Md Abdus and Al Mamoon, Ishtiak},
  journal={Data in Brief},
  volume={61},
  pages={111768},
  year={2025},
  publisher={Elsevier}
}

@article{edmund2017review,
  title={A review of substitute CT generation for MRI-only radiation therapy},
  author={Edmund, Jens M and Nyholm, Tufve},
  journal={Radiation Oncology},
  volume={12},
  number={1},
  pages={28},
  year={2017},
  publisher={Springer}
}

@article{bahloul2024advancements,
  title={Advancements in synthetic CT generation from MRI: A review of techniques, and trends in radiation therapy planning},
  author={Bahloul, Mohamed A and Jabeen, Saima and Benoumhani, Sara and Alsaleh, Habib Abdulmohsen and Belkhatir, Zehor and Al-Wabil, Areej},
  journal={Journal of Applied Clinical Medical Physics},
  volume={25},
  number={11},
  pages={e14499},
  year={2024},
  publisher={Wiley Online Library}
}

@article{dayarathna2024deep,
  title={Deep learning based synthesis of MRI, CT and PET: Review and analysis},
  author={Dayarathna, Sanuwani and Islam, Kh Tohidul and Uribe, Sergio and Yang, Guang and Hayat, Munawar and Chen, Zhaolin},
  journal={Medical image analysis},
  volume={92},
  pages={103046},
  year={2024},
  publisher={Elsevier}
}

@article{ibrahim2025generative,
  title={Generative AI for synthetic data across multiple medical modalities: A systematic review of recent developments and challenges},
  author={Ibrahim, Mahmoud and Al Khalil, Yasmina and Amirrajab, Sina and Sun, Chang and Breeuwer, Marcel and Pluim, Josien and Elen, Bart and Ertaylan, G{\"o}khan and Dumontier, Michel},
  journal={Computers in biology and medicine},
  volume={189},
  pages={109834},
  year={2025},
  publisher={Elsevier}
}

@article{iyer2025mri,
  title={MRI-to-CT Synthesis With Cranial Suture Segmentations Using A Variational Autoencoder Framework},
  author={Iyer, Krithika and Tapp, Austin and Paulli, Athelia and Dickerson, Gabrielle and Anwar, Syed Muhammad and Lepore, Natasha and Linguraru, Marius George},
  journal={arXiv preprint arXiv:2512.23894},
  year={2025}
}

@article{Li_2025,
doi = {10.1088/2632-2153/adc656},
url = {https://doi.org/10.1088/2632-2153/adc656},
year = {2025},
month = {apr},
publisher = {IOP Publishing},
volume = {6},
number = {2},
pages = {025007},
author = {Li, Yunxiang and Shao, Hua-Chieh and Qian, Xiaoxue and Zhang, You},
title = {FDDM: unsupervised medical image translation with a frequency-decoupled diffusion model},
journal = {Machine Learning: Science and Technology}
}

@article{guerreiro2017evaluation,
  title={Evaluation of a multi-atlas CT synthesis approach for MRI-only radiotherapy treatment planning},
  author={Guerreiro, F and Burgos, Ninon and Dunlop, A and Wong, K and Petkar, I and Nutting, C and Harrington, K and Bhide, S and Newbold, K and Dearnaley, D and others},
  journal={Physica Medica},
  volume={35},
  pages={7--17},
  year={2017},
  publisher={Elsevier}
}

@article{uh2014mri,
  title={MRI-based treatment planning with pseudo CT generated through atlas registration},
  author={Uh, Jinsoo and Merchant, Thomas E and Li, Yimei and Li, Xingyu and Hua, Chiaho},
  journal={Medical physics},
  volume={41},
  number={5},
  pages={051711},
  year={2014},
  publisher={Wiley Online Library}
}

@article{burgos2015ct,
  title={CT synthesis in the head \& neck region for PET/MR attenuation correction: an iterative multi-atlas approach},
  author={Burgos, Ninon and Cardoso, M Jorge and Modat, Marc and Punwani, Shonit and Atkinson, David and Arridge, Simon R and Hutton, Brian F and Ourselin, S{\'e}bastien},
  journal={EJNMMI physics},
  volume={2},
  number={Suppl 1},
  pages={A31},
  year={2015},
  publisher={Springer}
}

@article{chen2017attenuation,
  title={Attenuation correction of PET/MR imaging},
  author={Chen, Yasheng and An, Hongyu},
  journal={Magnetic Resonance Imaging Clinics},
  volume={25},
  number={2},
  pages={245--255},
  year={2017},
  publisher={Elsevier}
}

@inproceedings{roy2014mr,
  title={MR to CT registration of brains using image synthesis},
  author={Roy, Snehashis and Carass, Aaron and Jog, Amod and Prince, Jerry L and Lee, Junghoon},
  booktitle={Proceedings of SPIE},
  volume={9034},
  pages={spie--org},
  year={2014}
}

@inproceedings{lee2017multi,
  title={Multi-atlas-based CT synthesis from conventional MRI with patch-based refinement for MRI-based radiotherapy planning},
  author={Lee, Junghoon and Carass, Aaron and Jog, Amod and Zhao, Can and Prince, Jerry L},
  booktitle={Medical Imaging 2017: Image Processing},
  volume={10133},
  pages={434--439},
  year={2017},
  organization={SPIE}
}

@article{lei2019mri,
  title={MRI-only based synthetic CT generation using dense cycle consistent generative adversarial networks},
  author={Lei, Yang and Harms, Joseph and Wang, Tonghe and Liu, Yingzi and Shu, Hui-Kuo and Jani, Ashesh B and Curran, Walter J and Mao, Hui and Liu, Tian and Yang, Xiaofeng},
  journal={Medical physics},
  volume={46},
  number={8},
  pages={3565--3581},
  year={2019},
  publisher={Wiley Online Library}
}

@article{skandarani2023gans,
  title={Gans for medical image synthesis: An empirical study},
  author={Skandarani, Youssef and Jodoin, Pierre-Marc and Lalande, Alain},
  journal={Journal of Imaging},
  volume={9},
  number={3},
  pages={69},
  year={2023},
  publisher={MDPI}
}

@article{wang2022magnetic,
  title={Magnetic resonance-based synthetic computed tomography using generative adversarial networks for intracranial tumor radiotherapy treatment planning},
  author={Wang, Chun-Chieh and Wu, Pei-Huan and Lin, Gigin and Huang, Yen-Ling and Lin, Yu-Chun and Chang, Yi-Peng and Weng, Jun-Cheng},
  journal={Journal of personalized medicine},
  volume={12},
  number={3},
  pages={361},
  year={2022},
  publisher={MDPI}
}

@article{hsu2022synthetic,
  title={Synthetic CT generation for MRI-guided adaptive radiotherapy in prostate cancer},
  author={Hsu, Shu-Hui and Han, Zhaohui and Leeman, Jonathan E and Hu, Yue-Houng and Mak, Raymond H and Sudhyadhom, Atchar},
  journal={Frontiers in Oncology},
  volume={12},
  pages={969463},
  year={2022},
  publisher={Frontiers Media SA}
}

@article{lemus2022dosimetric,
  title={Dosimetric assessment of patient dose calculation on a deep learning-based synthesized computed tomography image for adaptive radiotherapy},
  author={Lemus, Olga M Dona and Wang, Yi-Fang and Li, Fiona and Jambawalikar, Sachin and Horowitz, David P and Xu, Yuanguang and Wuu, Cheng-Shie},
  journal={Journal of Applied Clinical Medical Physics},
  volume={23},
  number={7},
  pages={e13595},
  year={2022},
  publisher={Wiley Online Library}
}

@article{yang2020unsupervised,
  title={Unsupervised MR-to-CT synthesis using structure-constrained CycleGAN},
  author={Yang, Heran and Sun, Jian and Carass, Aaron and Zhao, Can and Lee, Junghoon and Prince, Jerry L and Xu, Zongben},
  journal={IEEE transactions on medical imaging},
  volume={39},
  number={12},
  pages={4249--4261},
  year={2020},
  publisher={IEEE}
}

@article{matsuo2022unsupervised,
  title={Unsupervised-learning-based method for chest MRI--CT transformation using structure constrained unsupervised generative attention networks},
  author={Matsuo, Hidetoshi and Nishio, Mizuho and Nogami, Munenobu and Zeng, Feibi and Kurimoto, Takako and Kaushik, Sandeep and Wiesinger, Florian and Kono, Atsushi K and Murakami, Takamichi},
  journal={Scientific reports},
  volume={12},
  number={1},
  pages={11090},
  year={2022},
  publisher={Nature Publishing Group UK London}
}

@article{gong2024channel,
  title={Channel-wise attention enhanced and structural similarity constrained cycleGAN for effective synthetic CT generation from head and neck MRI images},
  author={Gong, Changfei and Huang, Yuling and Luo, Mingming and Cao, Shunxiang and Gong, Xiaochang and Ding, Shenggou and Yuan, Xingxing and Zheng, Wenheng and Zhang, Yun},
  journal={Radiation Oncology},
  volume={19},
  number={1},
  pages={37},
  year={2024},
  publisher={Springer}
}

@article{liu2021ct,
  title={CT synthesis from MRI using multi-cycle GAN for head-and-neck radiation therapy},
  author={Liu, Yanxia and Chen, Anni and Shi, Hongyu and Huang, Sijuan and Zheng, Wanjia and Liu, Zhiqiang and Zhang, Qin and Yang, Xin},
  journal={Computerized medical imaging and graphics},
  volume={91},
  pages={101953},
  year={2021},
  publisher={Elsevier}
}

@article{pan2024synthetic,
  title={Synthetic CT generation from MRI using 3D transformer-based denoising diffusion model},
  author={Pan, Shaoyan and Abouei, Elham and Wynne, Jacob and Chang, Chih-Wei and Wang, Tonghe and Qiu, Richard LJ and Li, Yuheng and Peng, Junbo and Roper, Justin and Patel, Pretesh and others},
  journal={Medical Physics},
  volume={51},
  number={4},
  pages={2538--2548},
  year={2024},
  publisher={Wiley Online Library}
}

@article{ozbey2023unsupervised,
  title={Unsupervised medical image translation with adversarial diffusion models},
  author={{\"O}zbey, Muzaffer and Dalmaz, Onat and Dar, Salman UH and Bedel, Hasan A and {\"O}zturk, {\c{S}}aban and G{\"u}ng{\"o}r, Alper and Cukur, Tolga},
  journal={IEEE Transactions on Medical Imaging},
  volume={42},
  number={12},
  pages={3524--3539},
  year={2023},
  publisher={IEEE}
}

@article{lyu2022conversion,
  title={Conversion between CT and MRI images using diffusion and score-matching models},
  author={Lyu, Qing and Wang, Ge},
  journal={arXiv preprint arXiv:2209.12104},
  year={2022}
}

@article{deng2026generative,
  title={Generative Modeling via Drifting},
  author={Deng, Mingyang and Li, He and Li, Tianhong and Du, Yilun and He, Kaiming},
  journal={arXiv preprint arXiv:2602.04770},
  year={2026}
}

@article{nyholm2018goldatlas,
  author  = {Nyholm, Tufve and Svensson, Stina and Andersson, Sebastian and Jonsson, Joakim and Sohlin, Maja and Gustafsson, Christian and Kjell{\'e}n, Elisabeth and Muren, Ludvig P. and Maase, Hans von der and Wang, Jinyi and Ceberg, Sofie and Gunnlaugsson, Adalsteinn},
  title   = {MR and CT Data with Multiobserver Delineations of Organs in the Pelvic Area -- Part of the Gold Atlas Project},
  journal = {Medical Physics},
  year    = {2018},
  volume  = {45},
  number  = {3},
  pages   = {1295--1300},
  doi     = {10.1002/mp.12748}
}

@article{thummerer2023synthrad,
  author  = {Thummerer, Adrian and van der Bijl, Erik and Galapon Jr, Arthur and Verhoeff, Joost J. C. and Langendijk, Johannes A. and Both, Stefan and van den Berg, Cornelis A. T. and Maspero, Matteo},
  title   = {SynthRAD2023 Grand Challenge dataset: Generating synthetic CT for radiotherapy},
  journal = {Medical Physics},
  year    = {2023},
  volume  = {50},
  number  = {7},
  pages   = {4664--4674},
  doi     = {10.1002/mp.16529}
}

@ARTICLE{2015arXiv150504597R,
       author = {{Ronneberger}, Olaf and {Fischer}, Philipp and {Brox}, Thomas},
        title = "{U-Net: Convolutional Networks for Biomedical Image Segmentation}",
      journal = {arXiv e-prints},
         year = 2015,
        month = may,
          eid = {arXiv:1505.04597},
        pages = {arXiv:1505.04597},
          doi = {10.48550/arXiv.1505.04597},
archivePrefix = {arXiv},
       eprint = {1505.04597},
 primaryClass = {cs.CV},
       adsurl = {https://ui.adsabs.harvard.edu/abs/2015arXiv150504597R}
}

@ARTICLE{2013arXiv13126114K,
       author = {{Kingma}, Diederik P and {Welling}, Max},
        title = "{Auto-Encoding Variational Bayes}",
      journal = {arXiv e-prints},
         year = 2013,
        month = dec,
          eid = {arXiv:1312.6114},
        pages = {arXiv:1312.6114},
          doi = {10.48550/arXiv.1312.6114},
archivePrefix = {arXiv},
       eprint = {1312.6114},
 primaryClass = {stat.ML},
       adsurl = {https://ui.adsabs.harvard.edu/abs/2013arXiv1312.6114K}
}

@inproceedings{gulrajani2017improved,
  title={Improved training of wasserstein gans},
  author={Gulrajani, Ishaan and Ahmed, Faruk and Arjovsky, Martin and Dumoulin, Vincent and Courville, Aaron C},
  booktitle={Advances in neural information processing systems},
  volume={30},
  year={2017}
}

@ARTICLE{2024ITRPM8788H,
       author = {{Hein}, Dennis and {Holmin}, Staffan and {Szczykutowicz}, Timothy and {Maltz}, Jonathan S. and {Danielsson}, Mats and {Wang}, Ge and {Persson}, Mats},
        title = "{PPFM: Image Denoising in Photon-Counting CT Using Single-Step Posterior Sampling Poisson Flow Generative Models}",
      journal = {IEEE Transactions on Radiation and Plasma Medical Sciences},
         year = 2024,
        month = jan,
       volume = {8},
       number = {7},
        pages = {788-799},
          doi = {10.1109/TRPMS.2024.3410092},
archivePrefix = {arXiv},
       eprint = {2312.09754},
 primaryClass = {eess.IV},
       adsurl = {https://ui.adsabs.harvard.edu/abs/2024ITRPM...8..788H}
}

@ARTICLE{2024arXiv240514802J,
       author = {{Jiang}, Hongxu and {Imran}, Muhammad and {Zhang}, Teng and {Zhou}, Yuyin and {Liang}, Muxuan and {Gong}, Kuang and {Shao}, Wei},
        title = "{Fast-DDPM: Fast Denoising Diffusion Probabilistic Models for Medical Image-to-Image Generation}",
      journal = {arXiv e-prints},
         year = 2024,
        month = may,
          eid = {arXiv:2405.14802},
        pages = {arXiv:2405.14802},
          doi = {10.48550/arXiv.2405.14802},
archivePrefix = {arXiv},
       eprint = {2405.14802},
 primaryClass = {eess.IV},
       adsurl = {https://ui.adsabs.harvard.edu/abs/2024arXiv240514802J}
}

@ARTICLE{2020arXiv201002502S,
       author = {{Song}, Jiaming and {Meng}, Chenlin and {Ermon}, Stefano},
        title = "{Denoising Diffusion Implicit Models}",
      journal = {arXiv e-prints},
         year = 2020,
        month = oct,
          eid = {arXiv:2010.02502},
        pages = {arXiv:2010.02502},
          doi = {10.48550/arXiv.2010.02502},
archivePrefix = {arXiv},
       eprint = {2010.02502},
 primaryClass = {cs.LG},
       adsurl = {https://ui.adsabs.harvard.edu/abs/2020arXiv201002502S}
}

@ARTICLE{2020arXiv200611239H,
       author = {{Ho}, Jonathan and {Jain}, Ajay and {Abbeel}, Pieter},
        title = "{Denoising Diffusion Probabilistic Models}",
      journal = {arXiv e-prints},
         year = 2020,
        month = jun,
          eid = {arXiv:2006.11239},
        pages = {arXiv:2006.11239},
          doi = {10.48550/arXiv.2006.11239},
archivePrefix = {arXiv},
       eprint = {2006.11239},
 primaryClass = {cs.LG},
       adsurl = {https://ui.adsabs.harvard.edu/abs/2020arXiv200611239H}
}

@inproceedings{chen2020simple,
  title={A simple framework for contrastive learning of visual representations},
  author={Chen, Ting and Kornblith, Simon and Norouzi, Mohammad and Hinton, Geoffrey},
  booktitle={International conference on machine learning},
  pages={1597--1607},
  year={2020},
  organization={PmLR}
}

\end{document}